\documentclass{naturefigs}
\usepackage{graphicx}

\title{Quantum to Classical Transition in a Single-Ion Laser}

\author{Fran\c{c}ois  Dubin$^{1,2}$, Carlos Russo$^{1}$, Helena G.~Barros$^{1,3}$, Andreas Stute$^{1,3}$,\\ Christoph Becher$^{1,*}$, Piet O.~Schmidt$^{1,+}$ \& Rainer Blatt$^{1,3}$}

\begin{document}

\maketitle

\begin{affiliations}
\item Institut f\"ur Experimentalphysik, Universit\"at Innsbruck, Technikerstr. 25, 6020 Innsbruck, Austria
\item ICFO-Institut de Ci\`{e}ncies Fot\`{o}niques, Mediterranean Technology Park, 08860 Castelldefels, Spain
\item Institut f\"ur Quantenoptik und Quanteninformation, \"Osterreichische Akademie der Wissenschaften,
  Otto-Hittmair-Platz 1, 6020 Innsbruck, Austria
\end{affiliations}

\date{\today}% It is always \today, today,

\begin{abstract}
Stimulated emission of photons from a large number of atoms into the mode of a strong light field is the principle mechanism for lasing in ``classical" lasers\cite{Schawlow:1958}. The onset of lasing is marked by a threshold which can be characterised by a sharp increase in photon flux as a function of external pumping strength\cite{Mandel:1995, Gardiner:2004}. The same is not necessarily true for the fundamental building block of a laser: a single trapped atom interacting with a single optical radiation mode\cite{Cohen-Tannoudji:1998a,Milonni:1994,Berman:1994}. It has been shown that such a ``quantum" laser can exhibit thresholdless lasing in the regime of strong coupling between atom and radiation field\cite{An:1994,McKeever:2003}. However, although theoretically predicted\cite{Mu:1992,Pellizzari:1994,Briegel:1996,Meyer:1997}, a threshold at the single-atom level could not be experimentally observed so far. Here, we demonstrate and characterise a single-atom laser with and without threshold behaviour by changing the strength of atom-light field coupling. We observe the establishment of a laser threshold  through the accumulation of photons in the optical mode even for a mean photon number substantially lower than for the classical case. Furthermore, self-quenching occurs for very strong external pumping and constitutes an intrinsic limitation of single-atom lasers \cite{Mu:1992,Meyer:1997}. Moreover, we find that the statistical properties of the emitted light can be adjusted for weak external pumping,  from the quantum to the classical domain. Our observations mark an important step towards fundamental understanding of laser operation in the few-atom limit\cite{Bjoerk:1994} including systems based on semiconductor quantum dots \cite{Strauf:2006} or molecules \cite{Noginov:2009}.
\end{abstract}

Conventional lasers reach their threshold when the rate at which photons are coherently fed into the cavity mode exceeds losses and stimulated emission dominates over spontaneous emission into the lasing mode\cite{Mandel:1995, Gardiner:2004}. The most prominent characteristic of this threshold is a steep increase of photon flux from the laser while increasing the strength of the external pump exciting the gain medium. At the same time intensity fluctuations are reduced\cite{Rice:1994,Meyer:1997,Briegel:1996,Jones:1999}. These characteristics are induced by the non-linear process of photon stimulated emission of photons into the lasing mode.
At the single-atom level, characteristics that strongly deviate from standard laser properties have been predicted\cite{Jin:1994,Rice:1994,Briegel:1996,Meyer:1997,Wiseman:1997}, in particular regarding the question of the existence of a threshold.
For strong atom-cavity coupling no threshold exists. In this regime, photons are coherently scattered into the cavity mode through stimulation by the vacuum field \cite{McKeever:2003}. More importantly, strong coupling introduces a photon-blockade, which suppresses cavity mode populations with more than one photon\cite{Birnbaum:2005}. As a consequence, photon-stimulated emission enabling the threshold is absent, the laser threshold disappears and non-classical radiation is emitted\cite{McKeever:2003,Boozer:2004,Mu:1992,Meyer:1997}. It was shown theoretically that a one-atom laser can exhibit a threshold\cite{Meyer:1997,Briegel:1996} in an intermediate coupling regime. Here we explore a single-ion device in this regime that exhibits a laser threshold for strong external pumping for which photon-stimulated processes start playing a role, owing to the non-zero population of cavity modes with more than one photon.

The schematic experimental setup of our ion trap cavity-QED system is shown in Fig. 1\textbf{a}. Using frequency and intensity stabilised lasers, we excite a single $^{40}$Ca$^+$ ion which is confined in a linear Paul trap at the centre of a near-concentric optical cavity\cite{Russo:2009}. The cavity provides a coherent coupling rate between the ion and a selected optical mode of $g/2\pi= 1.3(1)$~MHz, limited by the localization of the ion along the cavity standing wave. With a cavity finesse of $\approx70 000$ at 866~nm, the field decay rate from the cavity is given by $\kappa$/2$\pi=54(1)$~kHz. A Hanbury-Brown\&Twiss setup consisting of a beamsplitter and two avalanche photodiodes is used to analyse photons leaving the cavity. It provides us with the steady-state population of photons in the cavity\cite{Russo:2009},  $n_\mathrm{ss}$, and allows us to compute the second order correlation function, $g^{(2)}(\tau)$ (see supplementary information).

In Fig. 1\textbf{b} we present a reduced three level picture of the $^{40}$Ca$^+$ ion's electronic structure which allows us to qualitatively describe the excitation process. Using an off-resonant drive laser, we achieve an effective coupling on a vacuum-stimulated Raman transition\cite{Cohen-Tannoudji:1998a} between the electronic ground state S$_{1/2}$ and the metastable state D$_{3/2}$. A transfer from S$_{1/2}$ to D$_{3/2}$ deposits a photon in the cavity.
The effective coupling between these levels is $g_{\mathrm{eff}}\sim g\cdot s_1$, where we have defined the drive parameter $s_1\equiv \Omega_1/2|\Delta_1|$, with $\Omega_1$ denoting the Rabi frequency of the 397~nm driving laser and $\Delta_1$ its fixed detuning from the S$_{1/2}$~--~P$_{1/2}$ resonance.
To close the excitation cycle, an infrared recycling laser at 866~nm, with a resonant Rabi frequency $\Omega_2$ and detuning $\Delta_2$, transfers population from the D$_{3/2}$ to the P$_{1/2}$ state. From there, spontaneous emission projects the ion back to the S$_{1/2}$ ground state at a rate $(2\gamma_1/2\pi)= 20~\mathrm{MHz}$.
The corresponding recycling rate is given by $2\gamma_\mathrm{r}=2\gamma_1 s_2^2$ with the recycling parameter $s_2\equiv\Omega_2/2\widetilde{\Delta}_2$. Here, we define the effective detuning $\widetilde{\Delta}_2\equiv\sqrt{\Delta_2^2+(\Omega_2/2)^2+\gamma^2}$ where $2\gamma=2(\gamma_1+\gamma_2)$ is the total decay rate from the P$_{1/2}$ state and 2$\gamma_2/2\pi=1.69$~MHz is the decay rate back to the D$_{3/2}$ state. The total cycling rate $\Gamma_\mathrm{tot}$ for external pumping, i.e. the rate at which one photon is added to the cavity mode, is thus given by $1/\Gamma_\mathrm{tot}\sim 1/2g_\mathrm{eff}+1/2\gamma_\mathrm{r}$. The strength of external pumping is then controlled by the drive and recycling lasers; however, note that the drive parameter $s_1$ is kept constant in our experiments where the external pumping strength is only varied by the recycling parameter. Furthermore, the  drive and recycling lasers introduce the broadenings $\gamma_{\mathrm{eff}}^{(1)}\sim \gamma s_1^2$ and $\gamma_{\mathrm{eff}}^{(2)}\sim\gamma s_2^2$, of the $S_{1/2}$ and $D_{3/2}$ states, respectively. The total linewidth of the $\mathrm{S}_{1/2}\leftrightarrow\mathrm{D}_{3/2}$ Raman transition is therefore given by
{$2\gamma_{\mathrm{eff}}=2(\gamma_{\mathrm{eff}}^{(1)}+\gamma_{\mathrm{eff}}^{(2)})=2\gamma(s_1^2+s_2^2)$}.
In this situation, an effective strong coupling is achieved for $g_{\mathrm{eff}}>(\gamma_{\mathrm{eff}},\kappa)$. In the following experiments, this condition is fulfilled for weak external pumping at a fixed detuning of the drive laser ($\Delta_1/2\pi\approx$ 350 MHz). Note that this laser also provides weak Doppler cooling to reduce the motion of the ion in the trap\cite{Russo:2009}.

We performed numerical simulations using a master equation approach to predict measured observables, i.e. the intra-cavity mean photon number, $n_\mathrm{ss}$, the Mandel $Q$-parameter\cite{Mandel:1995} and the ratio $R$ between coherent and incoherent emission rates on the lasing transition for continuous application of all lasers. The $Q$-parameter, $Q =n_\mathrm{ss}(g^{(2)}(0)-1)$, quantifies intensity fluctuations in the quantum ($Q<$ 0) and classical ($Q\geq$ 0) domains. The $R$-parameter is defined as  $R=\kappa\cdot n_\mathrm{ss}/(\gamma_2\cdot n_{P})$, where  $n_{P}$ is the steady-state population of the P$_{1/2}$ level, which controls the amplitude of incoherent photon scattering. Experimentally, $R$ is accessed by monitoring the fluorescence rate on the $\mathrm{S}_{1/2}\leftrightarrow\mathrm{P}_{1/2}$ transition which leads to $n_P$.

%===============================================================
%        Laser without threshold
%===============================================================

In the following, we show that the ion-cavity device acts as a thresholdless laser when operated at the onset of the strong coupling regime, i.e. for $g_\mathrm{eff}\sim\gamma_\mathrm{eff}^{(1)}$.
Raman resonance between selected S$_{1/2}$ and D$_{3/2}$ states is established at low recycling rates by adjusting the drive laser detuning\cite{Russo:2009}. In this situation, the strength of external pumping is given by the recycling laser parameter, $s_2$, which is held at a fixed detuning $\Delta_2/(2\pi)= -20 $~MHz.

Fig. 2 shows the measured steady-state photon number $n_\mathrm{ss}$ and the $Q$- and $R$-parameters as a function of external pumping together with simulation results (see supplementary information) for continuous laser excitation of the trapped ion at the boundary of the strong coupling regime ($g_\mathrm{eff}\sim\gamma_\mathrm{eff}^{(1)}$).
The following three regions are identified:
At low recycling parameters (region (I) of Fig. 2), we observe the signature of a thresholdless single-atom laser. It is marked by an increase of the intra-cavity mean photon number, $n_\mathrm{ss}$,  while intensity fluctuations, as indicated by the $Q$-parameter, are reduced. Lasing occurs in the quantum domain ($Q=-2.3(4)\%$ for $\Omega_2/(2\pi) \approx 4.6$~MHz) since single photons are stimulated into the cavity mode via the vacuum field while spontaneous emission is reduced $(R\approx 0.7\dots 0.8)$.
In region (II) of Fig. 2, the laser starts losing its quantum properties as it approaches the classical domain ($Q\geq 0$), while the intra-cavity mean photon number reaches a maximum.
For even stronger pumping, the recycling laser introduces phase noise leading to an increase of the $Q$-parameter (region (III) of Fig. 2).
At the same time it imposes a Stark shift of the D$_{3/2}$ levels, effectively reducing the atom-cavity coupling and therefore the mean photon number. This is a manifestation of the theoretically predicted self-quenching of single-atom lasers\cite{Mu:1992,Meyer:1997}.

%===============================================================
%        Tunable Statistics
%===========================================================
In Fig. 3\textbf{a-c}, we have investigated the dramatic change in the statistics of the emitted light observed in regions (I)-(III) of Fig. 2 in more detail for a slightly different (albeit compatible) set of experimental parameters.
Fig. 3\textbf{a} shows the measured and calculated normalized correlation functions $g^{(2)}(\tau)$ for a small total cycling rate $\Gamma_\mathrm{tot}/2\pi\sim 100~\mathrm{kHz}$. The slow recycling process introduces a time gap between successive photon emissions and sub-Poissonian light ($Q\sim -1.2(1)~\%$) is emitted\cite{Briegel:1996,Briegel:1996a}. Different timescales are apparent in Fig. 3\textbf{a}:
The decrease of the correlation function for $|\tau|< 1.5~\mu$s reveals the finite storage time for cavity photons. For longer times $\tau$, the $g^{(2)}$-function approaches unity on a timescale given by the total cycling rate $\Gamma_\mathrm{tot}$. Finally, for long time intervals $\tau$, a modulation of the correlation function is observed which can be attributed to the beat note between the ion's two quasi-degenerate radial modes of motion in the trap\cite{Rotter:2008}.
As we slightly increase the recycling parameter, the photon flux remains almost constant, but the statistical properties of the emitted photons change dramatically: The correlation function becomes almost flat [$Q=0.0(1)$~\% in Fig. 3\textbf{b}], akin to a Poissonian source of light. At even higher recycling parameters, the output statistics resembles that of a thermal source [$Q=+1.0(1)$~\% in Fig. 3\textbf{c}]. In general, the measured coincident photon correlations,  $g^{(2)}(0)$, are governed by the population of the cavity Fock states with two photons, $p_2$, so that $g^{(2)}(0)\sim 2p_2/p_1^2$, where $p_1$ is the population of states with only one photon. The population $p_2$ is established through an interplay between the rate $\Gamma_\mathrm{tot}$ at which photons are added to the cavity and the rate $2\kappa$ at which the photon state leaves the cavity (see supplementary Table 2). The tunability of this population $p_2$ and thus the statistical properties of the emitted light is a feature of our system, which can be adjusted to operate in the strong or weak coupling regimes, as in the following where we demonstrate a single-atom laser exhibiting threshold behavior.

%===============================================================
%        Laser with threshold
%===============================================================

In an experiment with different parameters, we studied the single-ion device for an even larger number of photons in the cavity. This was achieved by increasing the drive laser intensity to $\Omega_1$= $2\pi\cdot 130$~MHz. Experimental results are displayed in Fig. 4 where we identify the following regimes: For low recycling rates, i.e. in region (I), the photon number and $Q$-parameter both rise indicating classical non-lasing behaviour. Increasing the external pumping further (region (II)), $n_\mathrm{ss}$ still grows while the $Q$-parameter has passed through a maximum and decreases. Indeed, we note a 17~\% increase of the intra-cavity photon number together with a reduction of 33~\% in intensity fluctuations. At the same time, the rate of coherent vs. incoherent emission on the lasing transition indicated by the ratio $R$ is on the order of unity. These are the signatures for the onset of a laser threshold\cite{Meyer:1997,Briegel:1996}. As for a conventional laser, the threshold is reached when the population of photons in the cavity exceeds a critical value for which photon stimulated emission compensates incoherent processes. Most notably, this critical value is less than one for our one-atom device. Furthermore, we note that the variation of $n_\mathrm{ss}$ at threshold is less pronounced than for classical lasers.  This is a consequence of the low photon number in the lasing mode, as theoretically predicted \cite{Mu:1992,Rice:1994,Briegel:1996,Meyer:1997,Jones:1999}, caused by a saturation of the rate of external pumping ($\Gamma_\mathrm{tot}$) in region (II). Additionally, we observe in region (III) quenching of the laser output. Unlike the experiments presented in Fig. 2, this is not due to a level Stark shift since the Raman detuning is chosen such that resonant coupling is established at threshold. Rather, numerical simulations indicate that it is due to the formation of dark states by coherent coupling via $\Omega_2$, which disables the recycling process. More precisely, the induced dressed states, generally composed of two D$_{3/2}$ magnetic sub states and one P$_{1/2}$ sub level, lose their P$_{1/2}$ contribution in region (III), where $\Omega_2$ is large. This renders the external pumping inefficient hence yielding self-quenching which is a striking difference compared to conventional lasers. In fact, it constitutes an intrinsic limitation of single-atom lasers. In Fig. 4, we also present theoretical simulations which agree qualitatively with the experimental data. We attribute the discrepancy to the non-zero temperature of the ion, which modifies the excitation spectrum for the parameter range of the experiments presented in Fig. 4 and is not included in the simulations (see supplementary information).

In conclusion, we have demonstrated that a laser threshold can be achieved at the single-atom level, although much less pronounced compared to classical lasers\cite{Mu:1992,Rice:1994,Briegel:1996,Meyer:1997,Jones:1999} as a consequence of the low photon number in the lasing mode. Another striking difference between one-atom and classical lasers is the ability to tune the photon emission statistics from quantum to classical. Such a light source offers perspectives to study novel types of spectroscopy, based on narrow band optical excitation with quantum/classical statistical distributions\cite{Drummond:2004}.
Our trapped ion laser is ideally suited to further investigate the transition between quantum and classical lasers through controlled addition of more and more ions interacting with the light field.

\begin{addendum}
\item[Supplementary Information]is linked to the online version of the paper at
www.nature.com/nphys.
\item We acknowledge fruitful discussions with T. Salzburger, H. Ritsch, H.-J. Briegel and P. Zoller. We thank E.S. Phillips for assistance during the early stages of the experiment. This work has been partially supported by the Austrian Science Fund (SFB 15), by the European Commission (QUEST network, HPRNCT-2000-00121, QUBITS network, IST-1999-13021, SCALA Integrated Project, Contract No. 015714), and by the "Institut f\"ur Quanteninformation GmbH". C. Russo acknowledges support from the Funda\c{c}\~{a}o para a Ci\^{e}ncia e a Tecnologia -- SFRH/BD/6208/2001 and A. Stute acknowledges support from the Studienstiftung des deutschen Volkes. P.O.~Schmidt acknowledges support through the Center for Quantum Engineering and Space Time Research (QUEST) at the University of Hannover.
    \item[Author Contributions]  F.D. and C.R. contributed equally to this work.
    \item[Author information] The authors declare no competing
financial interests. Correspondence and requests for materials should be
addressed to P.O.S. (Piet.Schmidt@uibk.ac.at).
\end{addendum}

\normalsize{$^{*}$Fachrichtung Technische Physik, Universit\"{a}t des Saarlandes, Postfach 151150, 66041 Saarbr\"{u}cken, Germany}\\
\normalsize{$^+$QUEST Institut f\"{u}r Experimentelle Quantenmetrologie, Physikalisch-Technische Bundesanstalt und Leibniz Universit\"{a}t Hannover, Bundesallee 100, 38116 Braunschweig, Germany}

%======================================%
%<<<<<<<<<<<< BIBLIOGRAPHY >>>>>>>>>>>>>%
%======================================%

\bibliography{biblio-most-recent}

%======================================%
%<<<<<<<<<<<<<< FIGURES >>>>>>>>>>>>>>>>%
%======================================%

\clearpage

\newpage
\centerline{\includegraphics[width=\textwidth]{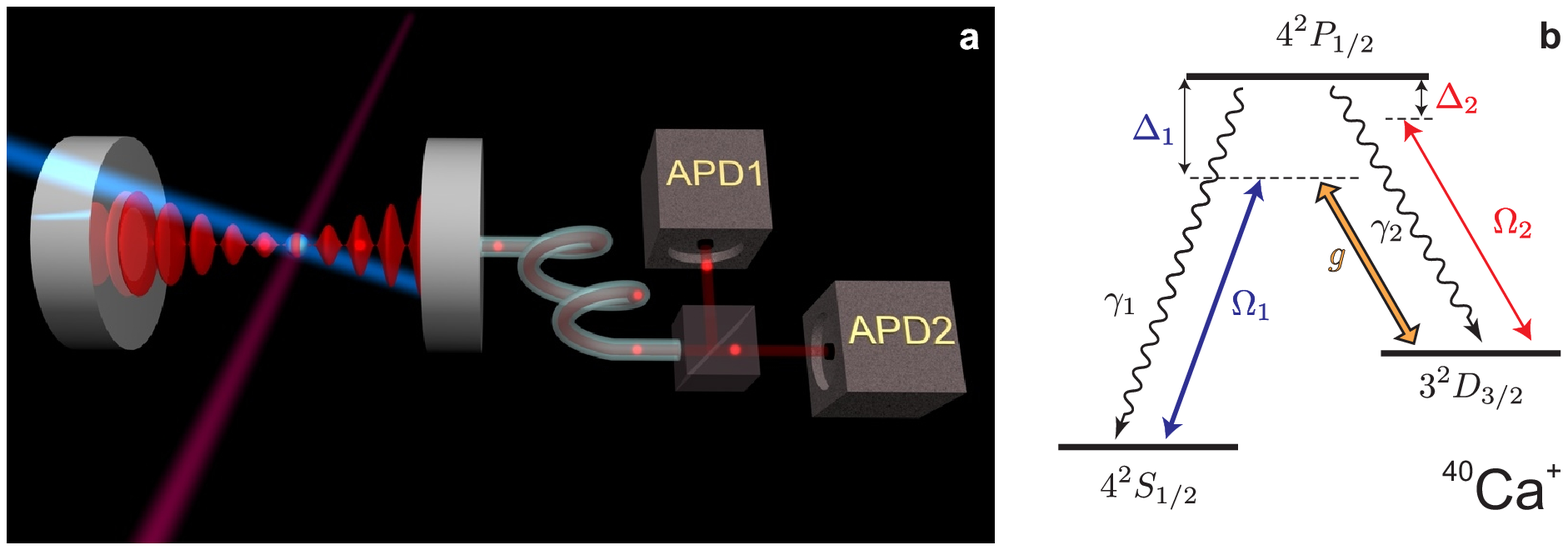}}
\noindent {\bf Figure 1.}  Experimental setup and excitation scheme. \textbf{a}, Schematic experimental setup: The ion (blue ball) is surrounded by a high-finesse cavity and positioned in an anti-node of the standing wave cavity field. The single-ion device is excited to a meta-stable state via a Raman transition (blue drive laser beam and cavity). A red laser provides repumping to close the excitation cycle. Photons leaving the cavity are coupled into a multi-mode optical fiber, split by a non-polarizing beam splitter and detected by two avalanche photodiodes (APD) in a Hanbury-Brown\&Twiss setup. Photon arrival times  are recorded with $\approx 300$~ps temporal resolution. \textbf{b}, Relevant level-scheme of the $^{40}$Ca$^+$ ion. A quantization axis is defined by applying a weak magnetic field, $B=2.8$~G, perpendicular to the cavity axis. The drive laser ($\Omega_1$) is polarized parallel to the \textbf{B} field: It excites transitions conserving angular momentum ($\Delta m=0$). The cavity (g) supports two linear polarizations, perpendicular and parallel to the \textbf{B} field. The recycling laser ($\Omega_2$) is polarized perpendicular to the \textbf{B} field.

\newpage
\centerline{\includegraphics[width=13cm]{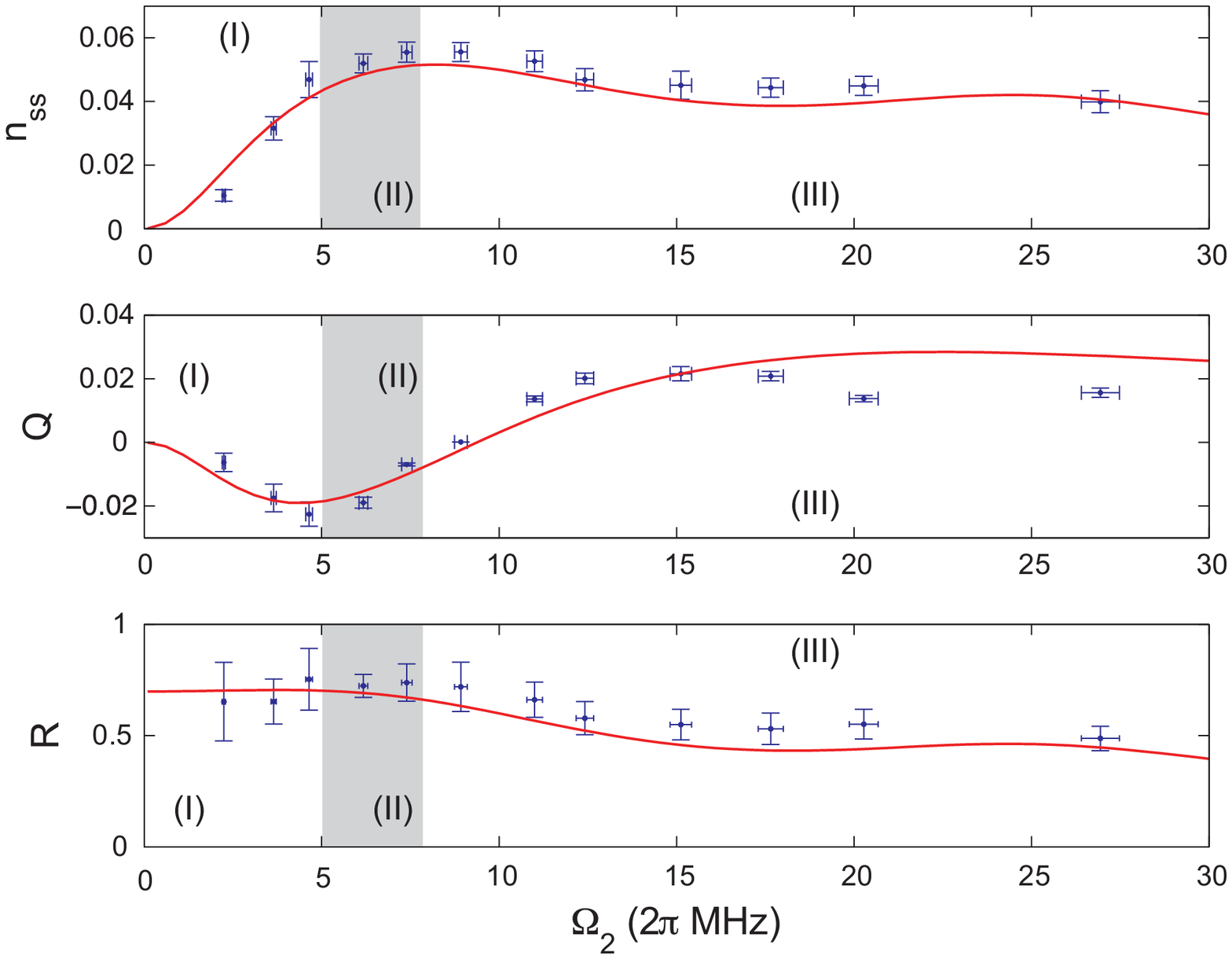} }
\noindent {\bf Figure 2.}  Thresholdless single-atom laser. Statistical behavior of the single-ion device at the boundary of the strong-coupling regime. Shown are the mean intra-cavity photon number, $n_\mathrm{ss}$, the Mandel $Q$ parameter and the ratio $R$ between coherent and incoherent scattering. Region (I) shows thresholdless lasing of the single-ion device, marked by decreasing intensity fluctuations ($\sim Q$) and increasing photon flux ($\sim n_\mathrm{ss}$) as a function of external pumping, while $R$ is on the order of one. In regions (II) and (III) the laser turns off (self-quenching). Drive laser parameters are ($\Omega_1$, $\Delta_1$)= $2\pi\cdot (88, -350)$~MHz leading to $g_\mathrm{eff}/(2\pi) = 165$~kHz $\sim$ $\gamma_{\mathrm{eff}}^{(1)}/(2\pi) = 170$~kHz. The recycling rate is varied by the Rabi frequency, $\Omega_2$, for a constant detuning $\Delta_2/2\pi\approx -20$~MHz. Experimental data are obtained with the same trapped ion continuously probed over 12 hours. Solid lines show the results of numerical simulations using independently calibrated experimental parameters. Error bars are derived from calibration uncertainties of the corresponding quantities (see supplementary information).
\newpage
\centerline{\includegraphics[width=8cm]{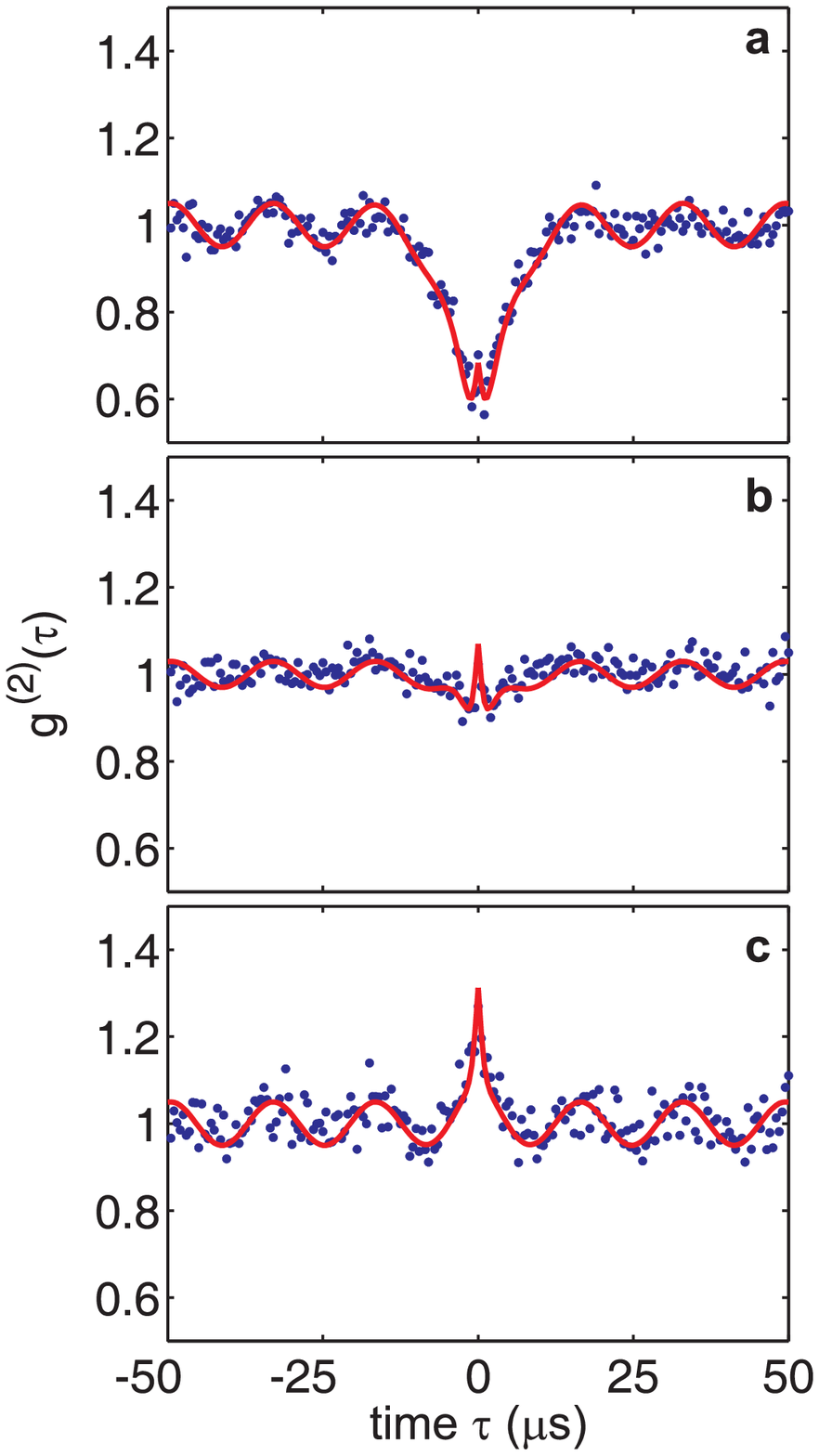}}
\noindent {\bf Figure 3.} Tunable photon statistics. Normalized second order photon correlations, $g^{(2)}(\tau)$ for different photon generation rates. Increasing the photon generation rate leads to a transition from sub-Poissonian (\textbf{a}) via almost Poissonian (\textbf{b}) to super-Poissonian, bunched (\textbf{c}) light emitted by the cavity. The peak observed around $\tau=0~\mu$s is a feature of the intermediate coupling regime and stems from the accumulation of photons inside the cavity. The drive laser excitation is such that $(\Omega_{1}, \Delta_1)\approx 2\pi (95, -400)$~MHz while the recycling laser is set at $(\Omega_2, \Delta_2)\approx 2\pi (7, -20)$~MHz in \textbf{a}, at $(\Omega_2, \Delta_2)\approx 2\pi (12, -20)$~MHz in \textbf{b}, and at $(\Omega_2, \Delta_2)\approx 2\pi (16, -20)$~MHz in \textbf{c}. In every measurement the cavity detuning is $\Delta_\mathrm{c}/2\pi\approx -400$~MHz,
the time resolution is set at 500~ns and accidental correlations have been subtracted (see supplementary information). Solid lines show the results of numerical simulations using independently calibrated experimental parameters.

\newpage

\centerline{ \includegraphics[width=14cm]{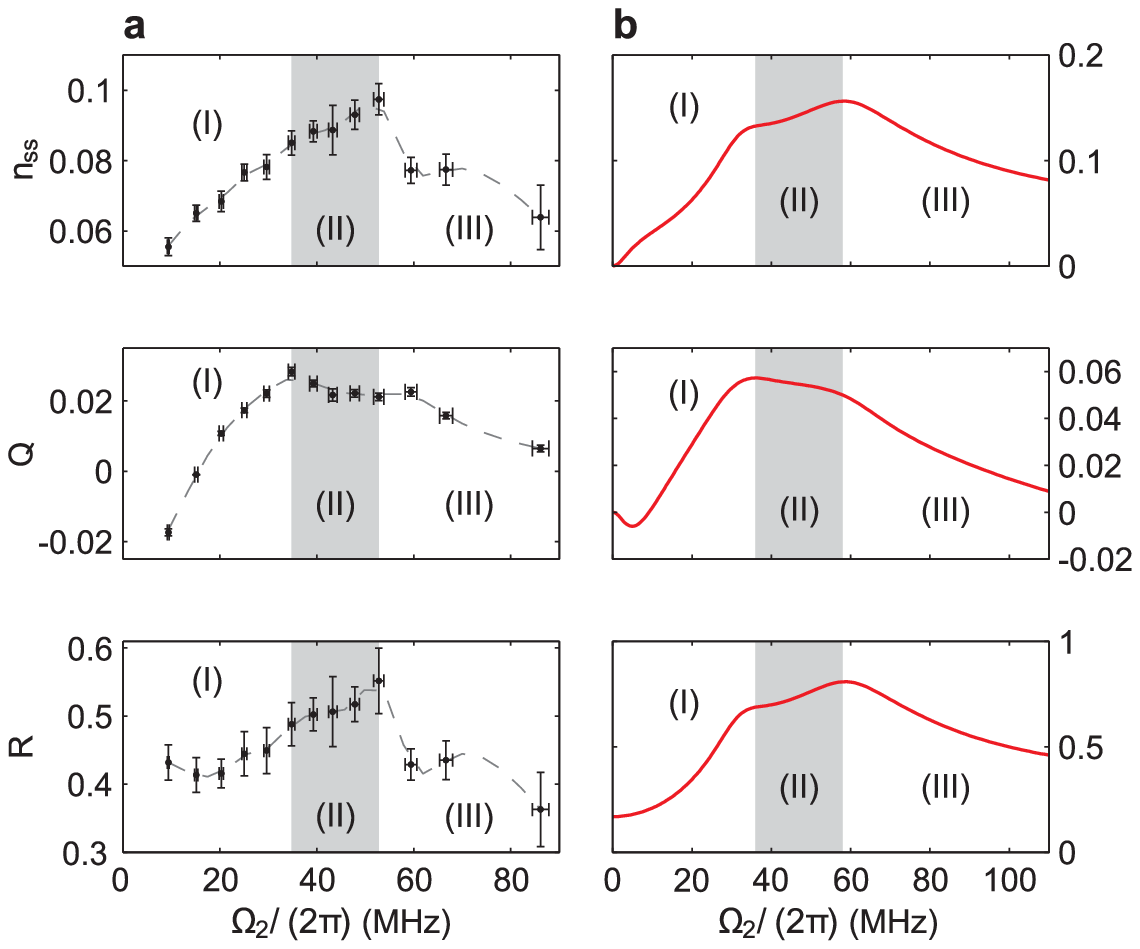} }
\noindent {\bf Figure 4.} Comparison between experimental observations  (\textbf{a}) and theoretical predictions (\textbf{b}) for the statistical response of the single-ion device operating as a laser with threshold behaviour.
Shown are the mean intra-cavity photon number, $n_\mathrm{ss}$, the Mandel $Q$ parameter and the ratio $R$ between coherent and incoherent scattering. Intensity ($\sim n_\mathrm{ss}$) and intensity fluctuations ($\sim Q$) rise as a function of external pumping in region (I), indicating classical behaviour. Signatures of a laser at threshold are observed in region (II) where $Q$ has passed through a maximum while $n_\mathrm{ss}$ is still rising and $R$ is on the order of one. Self-quenching turns the laser off in region (III). Drive laser parameters are ($\Omega_1$, $\Delta_1$)= $2\pi\cdot$ (130, -350) MHz, resulting in a $g_\mathrm{eff}\sim 240~$kHz. The recycling laser has a fixed detuning $\Delta_2/2\pi\approx -20$~MHz. Experimental data are obtained with the same trapped ion continuously probed over 12 hours. The dashed lines are a guide to the eye. Error bars are derived from calibration uncertainties of the corresponding quantities (see supplementary information).
\end{document}

\maketitle

\begin{affiliations}
\item Institut f\"ur Experimentalphysik, Universit\"at Innsbruck, Technikerstr. 25, 6020 Innsbruck, Austria
\item ICFO-Institut de Ci\`{e}ncies Fot\`{o}niques, Mediterranean Technology Park, 08860 Castelldefels, Spain
\item Institut f\"ur Quantenoptik und Quanteninformation, \"Osterreichische Akademie der Wissenschaften,
  Otto-Hittmair-Platz 1, 6020 Innsbruck, Austria
\end{affiliations}

\date{\today}% It is always \today, today,

%======================================%
% <<<<<<<<<<<< TITLE PAGE >>>>>>>>>>>>>>  %
%======================================%

%****************************************************************
\section{Calibration procedures}
%****************************************************************
\subsection{Cavity:}
%****************************************************************

The cavity decay rate was determined using a ring-down measurement\cite{Saleh:1991} with an error of less than 5~\%. We measured the frequency spacing between the TEM$_{00}$ and TEM$_{01}$ modes\cite{Siegman:1986} to obtain the exact cavity length. This allows us to calculate the cavity mode volume, and thus the ideal ion-cavity coupling rate, $g/2\pi= 1.61(2)$ MHz\cite{Berman:1994}.
However, we estimate that the ion-cavity coupling reduces to $g/2\pi= 1.3(1)$ MHz due to residual motion of the ion with respect to the trap\cite{Russo:2009}. In fact, our experiments are carried out at a large detuning of the drive laser, $\Delta_1/2\pi\approx$ 350 MHz, for which laser cooling has a reduced efficiency.

The length of the cavity is stabilized by servoing a piezo-electric transducer (PZT) attached to one of the cavity mirrors. The stabilization signal is derived using a transfer-lock scheme involving a frequency-stabilized laser at a wavelength of 785~nm which is resonant with one of the cavity modes. We estimate the stability of the cavity resonance frequency to be on the order of a few hundred kHz on a timescale of milliseconds. The instantaneous linewidth is expected to be close to the natural linewidth of the cavity. The relative position between the ion and the cavity can be adjusted by biasing the PZT attached to the second mirror of the cavity.

We observe the cavity excitation spectrum by applying a constant (Rabi frequency and detuning) drive laser while sweeping the cavity detuning across the Raman resonance by tuning the 785~nm laser frequency. Choosing a specific Raman resonance and fixed laser intensities and detunings, the cavity standing-wave can be mapped by recording the intra-cavity photon number via photon leakage through one of the cavity mirrors\cite{Guthohrlein:2001,Russo:2009} as a function of the bias voltage on the second mirror of the cavity.
From the visibility of the standing wave, the localization of the ion relative to the cavity can be inferred to be 100~nm\cite{Eschner:2003,Russo:2009}.
Delocalization is attributed to two effects with approximately equal contributions: (i) residual motion of the ion in the trap and (ii) mechanical vibration of cavity with respect to the trap.
%****************************************************************
\subsection{Magnetic field and trap frequencies:}
%****************************************************************
A bias magnetic field is induced by a pair of near-Helmholtz coils.
The magnitude of the magnetic field was first measured by resolving the frequency of several Zeeman lines on the quadrupole transition $\mathrm{S}_{1/2}\leftrightarrow \mathrm{D}_{5/2}$ at 729~nm\cite{Nagerl:2000}. This procedure was repeated for different currents, establishing a calibration curve with less than 5~\% error.
Additionally, the trap frequencies were determined by identifying the motional sidebands on the Zeeman lines\cite{Nagerl:2000}.

%****************************************************************
\subsection{Laser parameters:}
%****************************************************************

The relevant laser parameters are calibrated in the absence of ion-cavity interactions.
We record a collection of excitation spectra of resonance fluorescence to determine and/or set the excitation laser parameters, ($\Omega_{1,2}$, $\Delta_{1,2}$) and the laser linewidths $\delta_{1,2}$ of the drive (index 1) and recycle (index 2) lasers.
For each spectrum, the light emitted by the ion on the $S_{1/2}$--$P_{1/2}$ transition is monitored while the recycling laser detuning, $\Delta_2$, is scanned.
For the calibrations we typically record between five and ten spectra for fixed $\Omega_1$ and $\Delta_1$ while $\Omega_2$ is varied between each spectrum.
Similar spectra are recorded for different settings of $\Omega_1$ and $\Delta_1$.
Fitting all acquired spectra to a numerical model containing all 8 atomic states provides us with a calibration of all relevant experimental laser parameters with uncertainties as shown in Supplementary Table 1.
Please note that all laser intensities are actively stabilized and the stability of our experimental conditions is tested by periodically (approximately every 30~min) probing the ion.
The position of the ion in the cavity standing wave as well as possible laser frequency drifts are then continuously optimized during our experiments.

%\begin{table}
\begin{center}
\renewcommand{\arraystretch}{1.2}
\begin{tabular}{|c|c|c|}
  \hline
  parameter & typical value [MHz] & relative error \\\hline\hline
  $\Delta_1$ & $350$ 		& $2.5\times 10^{-3}$\\
  $\Omega_1$ & $0\dots 250$ 	& 5\% \\
  $\delta_1$ & $0.03$ 		& 12~\% \\
  $\Delta_2$ & $-20\dots 20$ 	& 10~\% \\
  $\Omega_2$ & $0\dots 80$ 	& 5~\% \\
  $\delta_2$ & $0.2$ 		& 70~\% \\
  $\Delta_c$ & $350$ 	& $2.5\times 10^{-3}$
  \\\hline
\end{tabular}
\end{center}
\noindent {\bf Supplementary Table 1:} Estimated uncertainties for the calibration of relevant experimental laser parameters. Note that the linewidth $\delta_2$ of the recycling laser has negligible influence on the experimental results.
%\end{table}

%****************************************************************
\subsection{Detection:}
%****************************************************************
We use two silicon avalanche photodiodes (PerkinElmer SPCM-AQR-15) with a measured quantum efficiency of $\approx 42\%$ at 866 nm and a specified dark count-rate of $<50$~counts/s in a Hanbury-Brown\&Twiss setup to detect photons emitted by the cavity (Figure 1). The arrival times of detected photons are then time-tagged (PicoQuant PicoHarp 300) with a sub-nanosecond time resolution. At steady-state, including all detection losses of our experimental setup, an intra-cavity mean photon number of one corresponds to a photo-electronic count-rate of 32(4)~kHz.

The photons spontaneously emitted on the $S_{1/2}$--$P_{1/2}$ transition are collected by an objective and are detected using a photomultiplier tube. The photon detection efficiency of this channel was determined from the fitting of a large series of excitation spectra of resonance fluorescence: $\eta_{397} = 1.8(1)\%$. Hence, the steady-state population of the P$_{1/2}$ levels is inferred, as well as the value of the $R$-parameter.

%****************************************************************
\section{Evaluation of the $g^{(2)}$ -- correlation function}
%****************************************************************
The photons emitted by the cavity are time-tagged with high efficiency by two fast photodiodes as described above.
The detection events are then grouped into 500~ns bins and the $g^{(2)}$-function, i.e. second-order time correlations among detection events, is calculated.
However, the accidental correlations must be subtracted.
These arise when background photons (stray light photons or APD dark counts) are correlated with themselves or output photons of the cavity.
To quantify the amount of accidental correlations we use the signal to noise ratio ($S/N$) from the cavity excitation spectrum measured together with the correlation function.
Considering that the noise follows a Poissonian time-distribution, the normalized correlation function can be calculated from the raw second-order correlation function of the detection events ($g^{(2)}_\mathrm{S+N}(\tau)$), and reads\cite{Keller:2004a}
\begin{equation}
g^{(2)}(\tau) = 1 + \frac{(S/N +1)^2}{(S/N)^2} \cdot \left(g^{(2)}_\mathrm{S+N}(\tau) - 1 \right).
\end{equation}

\section{Theoretical model}

\subsection{General description:}

Our theoretical predictions are based on a master equation approach for the total density matrix of the coupled ion-cavity system, $\rho$. The latter is expanded including eight relevant electronic levels for the ion part and two degenerate cavity modes with crossed polarization ($h$ and $v$), in a number basis with at most four Fock states for each mode. The laser excitation of the ion is treated semi-classically whereas the ion-cavity interaction is written in the Jaynes-Cummings form in the rotating wave approximation\cite{Gardiner:2004}. All decays, losses and laser linewidths have corresponding Liouvillians entering the master equation expressed in the Lindblad form.
The time evolution of the density matrix, $\rho(t)$, and the steady-state $\rho_{ss}$ are hence obtained by numerically solving a master equation for our experimental situation.

\subsection{Steady-state photon number:}

The steady-state intra-cavity photon number is directly obtained from $\rho_\mathrm{ss}$. It reads  $n_\mathrm{ss}$=$\langle n \rangle_\mathrm{ss}$, where $n=(a_h^\dag a_h+a_v^\dag a_v)$, and $a_h^\dagger$ and $a_v^\dagger$ are the creation operators for cavity photons with horizontal (along the atomic quantization axis) and vertical polarization, respectively. Here we have defined the expectation value of the photon number, $n$, with respect to the steady-state density matrix, i.e. $\langle n \rangle_\mathrm{ss}$=Tr[$n\rho_\mathrm{ss}$], Tr[...] denoting the evaluation of the trace along the ion-cavity subspace.

\subsection{Second-order correlation function $g^{(2)}(\tau)$ and Mandel $Q$-parameter:}

The second-order correlation function, $g^{(2)}(t,t+\tau)$, gives the probability to detect a photon at a time ($t+\tau$) conditioned on a previous photon detection at time $t$. It is usually referred to as intensity correlation function and reads
\begin{equation}
g^{(2)}(t,t+\tau)=\frac{\langle:I(t)I(t+\tau):\rangle}{|\langle I(t)\rangle||\langle I(t+\tau)\rangle|},\label{g2-def}
\end{equation}
where  the intensity-operator, $I(t)$, is proportional to the intra-cavity photon number, $I(t)\propto n(t)$. Furthermore, note that operators inside colons must be time and normally ordered in Supplementary Equation (\ref{g2-def}). At steady-state, i.e. for $t\rightarrow\infty$, the normalized correlation-function is a function of $\tau$ only and it follows
\begin{equation}
g^{(2)}(\tau)=\frac{\langle n(\tau)\rangle_{0}}{n_{ss}}.\label{norm-g2}
\end{equation}
In Supplementary Equation (\ref{norm-g2}), the expectation value $\langle...\rangle_0$ is evaluated using the conditional density matrix $\rho_0=(a_h\rho_{\mathrm{ss}}a_h^{\dag}+a_v\rho_{\mathrm{ss}}a_v^{\dag})/n_{ss}$. It shows that the $g^{(2)}$-function resolves the transient evolution of the intra-cavity photon number, upon annihilation of a photon at steady state. It directly relates to intensity fluctuations via the Mandel $Q$-parameter which reads $Q$=$n_\mathrm{ss}(g^{(2)}(0)-1)$=$\langle n\rangle_{0}-n_\mathrm{ss}$.

Finally, let us mention that our theoretical model allows us to evaluate any observable of the single-ion device. For instance, we confirm in the following that the measured correlations for low recycling rates $g^{(2)}(0)\approx 2p_2/p_1^2$ (Figure 3) by extracting the populations $p_1$ and $p_2$ for the one- and two-photon Fock state, respectively.  Supplementary Table 2 compares our simulations with the experimentally obtained $g^{(2)}_\mathrm{exp}(0)$.

 \begin{center}
 \renewcommand{\arraystretch}{1.2}
 \begin{tabular}{|c|c|c|c|c|}
   \hline
   Fig. & $p_1$ & $p_2$ & $2p_2/p_1^2$ & $g^{(2)}_\mathrm{exp}(0)$ \\
   \hline\hline
   3(A) & $3.5\times 10^{-2}$ & $4.1\times 10^{-4}$ & 0.67 & 0.70 \\
   3(B) & $4.2\times 10^{-2}$ & $9.7\times 10^{-4}$ & 1.09 & 1.02 \\
   3(C) & $4.4\times 10^{-2}$ & $13 \times 10^{-4}$ & 1.34 & 1.27 \\
 \hline
 \end{tabular}
 \end{center}
\noindent {\bf Supplementary Table 2:} Simulated one- and two-photon Fock state populations. This leads to $g^{(2)}(0)\approx 2p_2/p_1^2$ which is compared to the experimentally observed $g^{(2)}_\mathrm{exp}(0)$.

\subsection{Influence of the motion of the ion:}

For the experiments reported in Figure 2, we reach a quantitative agreement between our observations and the results of our theoretical simulations solely considering independently measured experimental parameters. By contrast, for stronger excitation of the drive laser, as in Figure 4, we achieve only qualitative agreement.

In Figure 4, theoretical simulations are compared to experimental findings when considering calibrated experimental parameters. Qualitatively, these exhibit the same statistical variations as a function of external pumping, but the measured intra-cavity mean photon number differs from the predictions. Consequently, the amplitudes of $Q$ and $R$ differ as well. We mainly attribute the observed discrepancy to the motional degrees of freedom of the ion which are not considered in our model. In the pumping process, the drive laser for the Raman transition provides only weak cooling, since it is far detuned from the S$_{1/2}$- P$_{1/2}$ resonance. This results in a non-negligible motional excitation of the ion in the trap. As a consequence, the atom-cavity coupling, g, is reduced by up to 20$\%$ and motional sidebands involving the absorption or emission of phonons complicate the observed Raman spectrum\cite{Russo:2009}.

In the experiments reported in Figure 4, the cavity mode is set to be on the slope of the Raman resonance at low recycling rates. Resonant coupling is then achieved in region (II) due to the Stark shift induced by the recycling laser. At the same time, the shape of the resonance and the resultant experimental parameters $(n_\mathrm{ss}, Q, R)$ are strongly distorted by the motional sidebands\cite{Russo:2009}. Thereby, quantitative agreement between simulation and experiment is lost. On the other hand, in the experiments presented in Figure 2, the single-ion device is operated close to the Raman resonance throughout the investigated parameter range. Hence, the ion motional degrees of freedom play a minor role and theoretical predictions quantitatively reproduce experimental findings.\\[0.5cm]

\section{Supplementary References}
\bibliography{biblio-most-recent}

\normalsize{$^{*}$Fachrichtung Technische Physik, Universit\"{a}t des Saarlandes, Postfach 151150, 66041 Saarbr\"{u}cken, Germany}\\
\normalsize{$^+$QUEST Institut f\"{u}r Experimentelle Quantenmetrologie, Physikalisch-Technische Bundesanstalt und Leibniz Universit\"{a}t Hannover, Bundesallee 100, 38116 Braunschweig, Germany}